\documentstyle[12pt,epsfig]{ioplppt}

\def\half{\frac{1}{2}}
\def\<{\langle}
\def\>{\rangle}
\def\identity{1\hspace{-0.4em}1}
\newcommand{\Title}[1]{}

\begin{document}
\jl{1}
\title{Continuous stochastic Schr\"odinger equations and localization}
\author{M. Rigo, F. Mota-Furtado and P.F. O'Mahony}
\address{Department of Mathematics, Royal Holloway, University of
London, Egham, Surrey TW20 0EX, UK}

\begin{abstract}
The set of continuous norm-preserving stochastic Schr\"odinger
equations associated with the Lindblad master equation is
introduced. This set is used to describe the localization properties
of the state vector toward eigenstates of the environment operator.
Particular focus is placed on determining the stochastic equation
which exhibits the highest rate of localization for wide open
systems. An equation having such a property is proposed in the case of
a single non-hermitian environment operator. This result is relevant
to numerical simulations of quantum trajectories where localization
properties are used to reduce the number of basis states needed to
represent the system state, and thereby increase the speed of
calculation.
\end{abstract}

\section{Introduction}

A quantum system interacting with its environment can be described, in
the Markovian approximation, by two complementary approaches. In the
first and most commonly used \cite{Louisell,Cohen,Gardiner2}, the
system is represented by a density operator and its evolution is
described by a master equation. In the second a state vector
represents the system and a stochastic Schr\"odinger equation
describes the state evolution \cite{PB,ZG,C93} These two treatments
are equivalent in the following sense: for all times an ensemble of
state vectors generated by a stochastic Schr\"odinger equation
reproduces, on average, the density operator generated by the master
equation. The correspondence is not uniquely defined, in that for a
single master equation there are many associated stochastic equations.

Stochastic state vector equations, also called {\it unravelings} of
the master equation, have been introduced in different contexts and
with different interpretations. In the fundamental theory of quantum
measurement, stochastic equations have been used to describe the
general dynamical process of the state collapse into an eigenstate of
the measured observable, i.e. {\it localization}
\cite{PB,G89,GP92,P94,D88,HZ95,Z96,SP}. In quantum optics, stochastic
Schr\"odinger equations have been used to describe the system state
conditioned by measurement outcomes. In this context, an unraveling
corresponds to a specified measurement scheme, such as photon
counting, heterodyne or homodyne detection \cite{C93,WM93a}.  More
generally, in the field of open quantum systems, unravelings have been
used as an efficient numerical method to solve the master equation
\cite{DCM92,GPZ92,MCD93,CM95,NS95,MC96,N96,HMMZ96,SAP95,SBP95,SBP96,SB97}.

The present work is motivated by a recent study of a quantum system in
interaction with a thermal bath using the quantum jump (QJ) unraveling
\cite{BGOR97}. It is shown that, under some assumptions valid in the
classical limit, the QJ trajectories, i.e., the realization of the
stochastic process, approach a diffusive limit very similar to the one
exhibited by the quantum state diffusion (QSD) trajectories. Since
diffusion is expected to be a general feature associated with the
emergence of classicality, a description of the whole set of
continuous unravelings becomes important. This set is introduced in a
unified way in the following section. We will show that each
continuous unraveling can be characterized very simply by specifying
its noise correlations. The set of continuous unravelings is then used
to study how quantum state properties, such as localization, evolve
when the unraveling changes.

A very important characteristic of quantum trajectories which has both
physical and numerical consequences is the localization of quantum
states toward eigenstates of the environment operator.  Working with
the real noise unraveling (RN), Gisin~\cite{G89} has shown that for an
arbitrary hermitian Lindblad operator $L$ the state vectors
concentrate on the eigenspace of $L$.  Percival~\cite{P94} extended
this result by giving a proper definition of the ensemble localization
of an arbitrary operator and then providing analytical bounds for the
rate of self-localization of hermitian and nonhermitian Lindblad
operators for the quantum state diffusion unraveling. For a
dissipative interaction, Garraway and Knight \cite{GK94a,GK94b} have
presented numerical simulations of the localization process using the
QJ and QSD unravelings. Starting from different quantum states, such
as a superposition of two coherent states, a Fock state and a squeezed
ground state, they have shown that such states are highly sensitive to
dissipation. They also illustrated the localization process. Recently
they have applied their results to describe the evolution of a
Schr\"odinger cat state in a Kerr medium where localization competes
with nonlinear effects \cite{GK96} (see also \cite{RAMFO97}).

For numerical simulation of open quantum systems, individual
trajectories have proven advantageous over density operator
computations. The main advantages stem from the fact that less space
is needed to store and propagate in time a state vector than a density
matrix. In addition, for trajectory methods one can exploit the
localization property to reduce the number of basis states needed to
represent the state vector, thus significantly reducing the time
needed to calculate quantum trajectories. For quantum jump
unravelings, when many Lindblad operators are present, it is well
known that one can perform a unitary transformation to select one of
the quantum jumps unravelings, in such a way as to minimize the number
of basis states needed (see ref.  \cite{HMMZ96} for an application of
this property). The localization of the state vector for QSD has been
exploited in the mixed representation of Steimle and al. \cite{SAP95}
and the moving basis of Schack and al.~\cite{SBP95,SBP96,SB97}.

In section \ref{SecLoc} we use the set of continuous unravelings to
describe localization properties for a single environment operator.
Several well known results are recovered for a hermitian operator
\cite{PB,GPPC}. In the case of a non-hermitian operator the minimal
rate of localization introduced for the QSD unraveling
\cite{P94,HZ95,Z96} is extended to the complete set showing that
localization is a general feature shared by all the continuous
unravelings, and a new unraveling is introduced. Some theoretical
arguments supported by numerical simulations suggest that this new
unraveling possesses the highest localization rate.

In the present work we make use of the freedom of choice for the noise
correlations to obtain the continuous unraveling which localizes the
state vector the most. This transformation should not be confused with
the unitary transformation discussed above. These two transformations
are complementary and can be used together. At the end of section
\ref{SecLoc} we compare the localization properties of QSD and of our
proposed unraveling.

Finally, in section \ref{SecDiscussion} we summarize our results and
draw conclusions about the applicability of the set of continuous
unravelings to the study of the quasi-classical limit of open quantum
systems.

\section{Continuous unravelings}
We proceed following closely the derivation of the quantum state
diffusion unraveling by Gisin and Percival \cite{GP92}.  In this work,
a general stochastic differential equation with a complex Wiener
process is used as a starting point. The drift and noise terms are
then specified by asking that the stochastic differential equation
recovers on average the Lindblad master equation for the density
operator $\rho$ of the system
\begin{equation}
{\dot \rho} = - \frac{i}{\hbar} [H,\rho] + \sum_{j=1}^J \left( L_j
\rho L_j^\dagger - {1\over2} \{ L_j^\dagger L_j, \rho \} \right),
\label{EqMaster}
\end{equation}
where $H$ is the system Hamiltonian and $L_j$ ($j=1,\ldots ,J$) the
set of Lindblad operators which represent the influence of the
environment. (Since the master equation (\ref{EqMaster}) is valid under
a Markovian approximation, all the stochastic differential equations
considered apply only within this approximation). The other conditions
needed to specify the drift and noise terms are that the state remains
normalized and that the stochastic equation shares the same invariance
properties under unitary transformations as the master equation. This
last constraint is used to prove the uniqueness of QSD among the set
of continuous unravelings. Removing the constraint of invariance under
unitary transformations among the Lindblad operators, we obtain the
set of continuous norm-preserving unravelings related to the master
equation.

\subsection{Derivation of the stochastic Schr\"odinger equations}
We start by considering a general stochastic differential equation of the
following It\^o form which gives the variation $|d\psi\>$ of the state
vector $|\psi\>$ in a time $dt$
\begin{equation}
|d\psi\> = |v\> dt +\sum_{j=1}^J|u_j\> 
\left(\sum_{n=1}^N\alpha_{jn}dW_{jn}\right)
\label{EqStocha}
\end{equation}
where $|v\>$ and $|u_j\>$ are vectors, $\alpha_{jn}$ are complex
numbers and $dW_{jn}$ independent real Wiener processes
\cite{Gardiner1} which obey the following relationships
\begin{equation}
M(dW_{jn})=0  ~~~~ dW_{jn}dW_{km}=\delta_{jk}\delta_{mn}dt
~~~~ dW_{jn} dt =0
\end{equation}
where $M$ represents the ensemble average.  The two conditions to be
respected by the previous equation (\ref{EqStocha}) are ($i$) the
state is normalized for all times $ \<\psi|\psi\>_t=1$ and ($ii$) for
each time, the mean of the projector associated to the state $|\psi\>$
gives the density matrix $\rho = M(|\psi\>\<\psi|)$ with the density
matrix $\rho$ evolving according to the master equation in Lindblad
form (\ref{EqMaster}).  In the following, these two conditions will be
used to relate the drift term $|v\>$ and the stochastic term $|u_j\>$
to the state $|\psi\>$ as well as giving constraints on the complex
numbers $\alpha_{jn}$. Notice that the $\alpha_{jn}$ may also depend
on the state $|\psi\>$ and on time $t$.

By following closely the QSD derivation given in reference
\cite{GP92}, we obtain the expression for the drift term
\begin{equation}
|v\> = -\frac{i}{\hbar} H|\psi\> 
- \half\sum_j \left( L_j^\dagger L_j +\<L_j^\dagger\>_\psi\< L_j\>_\psi 
- 2\<L_j^\dagger\>_\psi L_j \right)|\psi\> 
\label{EqDriftTerm}
\end{equation}
where $\<L_j\>_\psi = \<\psi| L_j|\psi\>$ is the expectation value of
$L_j$ for the state $|\psi\>$. The drift term $|v\>$ is the same as
that obtained in the QSD derivation, but the stochastic vectors
$|u_j\>$ become
\begin{equation}
|u_j\> =\frac{1}{\sqrt{\sum_n|\alpha_{jn}|^2}}
\sum_{k} \beta_{jk}(L_k-\<L_k\>_\psi)|\psi\> ~~~~j=1\ldots J
\label{EqNoiseTerm}
\end{equation}
which differs from the QSD derivation by the introduction of the
normalization factor $(\sum_n|\alpha_{jn}|^2)^{-1/2}$ and the set of
complex numbers $\beta_{jk}$. The latter are arbitrary coefficients of
a $J\times J$ unitary matrix which arises due to the freedom of choice
in the linear combination of vectors $(L_k-\<L_k\>_\psi)|\psi\>$ used
to express $|u_j\>$.

Finally one gets the equation for the state vector increment
\begin{eqnarray}
|d\psi\> &=& -\frac{i}{\hbar} H|\psi\> dt 
- \half\sum_{j=1}^J \left( L_j^\dagger L_j +\<L_j^\dagger\>_\psi\< L_j\>_\psi 
- 2\<L_j^\dagger\>_\psi L_j \right)|\psi\> dt \nonumber \\ && 
+ \sum_{k=1}^J \left( L_k -\< L_k\>_\psi \right)|\psi\> d\zeta_k 
\end{eqnarray}
This equation shows that all the indeterminacy due to the coefficients
$\alpha_{jn}$ and to the unitary transformation $(\beta_{jk})$ can be
included in the noise terms $d\zeta_j$ which are given by
\begin{equation}
d\zeta_k = \sum_{j=1}^J\beta_{jk}\frac{\sum_n \alpha_{jn}dW_{jn}}{\sqrt{\sum_n|\alpha_{jn}|^2}}
\label{EqNoise1}
\end{equation}
It can be seen easily that they have zero mean $M(d\zeta_j)=0$ and
correlations
\begin{equation}
d\zeta_jd\zeta_k^\ast = \delta_{jk} dt
~~~~ \mbox{and} ~~~~ 
d\zeta_jd\zeta_k = c_{jk} dt 
\label{EqNoise2}
\end{equation}
where $c_{jk}$ are correlation coefficients related to the unitary
transformation $(\beta_{jk})$ and the noise coefficients $\alpha_{in}$
in the following way
\begin{equation}
c_{jk} = \sum_{i=1}^J \beta_{ij}\beta_{ik} c_i
~~~ \mbox{with} ~~~c_i = \frac{\sum_n \alpha_{in}^2}{\sum_n|\alpha_{in}|^2} 
\label{EqNoise3}
\end{equation}
and $c_i$ are complex numbers inside the unit circle $|c_i|\leq 1$ for
all $i=1, \ldots, J$ (see Appendix). Notice that although the
stochastic process is completely specified by the numbers $\beta_{jk}$
and $\alpha_{in}$, that specification will be non-unique in that all
sets of numbers that yield the same correlation coefficients $c_{jk}$
through equation (\ref{EqNoise3}) will describe the same stochastic
process \cite{Gardiner1}. Thus an unraveling is completely specified
when the correlation coefficients $c_{jk}$ are given. This result
provides a natural classification for continuous stochastic
Schr\"odinger equations associated with the completely positive master
equation (\ref{EqMaster}). In ref \cite{G90} a similar classification
is given for all (Markovian) positive master equations but only for
two dimensional Hilbert spaces. This suggests that the same
classification procedure can be extended to every positive master
equation in arbitrary dimensions.

In the absence of a unitary transformation $(\beta_{jk}=\delta_{jk})$
and with only one Wiener process $N=1$, the complex noises $d\zeta_j$
are necessarily of the form $d\zeta_j=e^{i\phi_j} dW_{j1}$. For
$\phi_j=0$ we recover the real noise unraveling and for $\phi_j=\pi/2$
the imaginary noise unraveling. The present work shows not only that
any phase $\phi_j$ can be chosen but also that the phase can be a
smooth function of the state $|\psi\>$ and of the time. The QSD
unraveling is recovered taking two Wiener processes ($N=2$) with
$d\zeta_j = (dW_{j1}+idW_{j2})/\sqrt{2}$ for all $j$. In this special
case the correlations $c_{jk}$ vanish.

\subsection{Unitary transformation}
Let us introduce the following unitary transformation among Lindblad
operators
\begin{equation}
L_j = \sum_k u_{jk} \tilde{L}_k -\lambda_j \identity
\label{EqLtransfo}
\end{equation}
where $u_{jk}$ and $\lambda_j$ are complex numbers and $(u_{jk})$ a
unitary matrix~\cite{GP92,P94,RG96}. With this transformation the
noise terms become $d\tilde{\zeta}_k = \sum_j u_{jk}d\zeta_j $ with
the correlations
\begin{equation}
d\tilde{\zeta}_j d\tilde{\zeta}_k^\ast=\delta_{jk}dt
  ~~~~\mbox{and}~~~~ d\tilde{\zeta}_j d\tilde{\zeta}_k=\sum_{m,n=1}^J
  u_{mj} u_{nk}c_{mn} dt
\end{equation}
These correlations will depend on the unitary transformation $(u_{jk})$
unless all the correlation factors vanish, i.e. $c_{jk}=0$. Since
$(\beta_{jk})$ is itself a unitary transformation, a necessary
condition for invariance under unitary transformation is given by
\begin{equation}
c_j=0 ~~~\mbox{for all} ~~~ j=1,\ldots,J.
\label{EqUnitInvCond}
\end{equation}
When only one Wiener process $N=1$ is present, the unitary invariance
condition (\ref{EqUnitInvCond}) implies $\alpha_{j1}=0$ for all $j$. As
a consequence there is no invariant unraveling with only one Wiener
process. With two Wiener processes $N=2$, the invariance condition
becomes $\alpha_{j1}^2+\alpha_{j2}^2=0$. The norm of the two complex
numbers is the same $|\alpha_{j1}|=|\alpha_{j2}|$ and the phases are
related by $\phi_{j1}-\phi_{j2}=\pi/2+n\pi$ where $n$ is any integer
number. The simplest case $n=0$ leads to
\begin{equation}
d\zeta_j = e^{i\phi_j}\left(\frac{dW_{j1}+idW_{j2}}{\sqrt{2}}\right)
\end{equation}
which correspond to the complex noise used in the QSD unraveling when
the phases $\phi_j$ are set to zero. The simplest case which can
satisfy the invariance condition (\ref{EqUnitInvCond}) is given by the
QSD unraveling. The phases $\phi_j$ and other choices of $n$ introduce
only irrelevant phase factors which can be neglected.  This is the
uniqueness result of Gisin and Percival for QSD.  If one considers
more than two Wiener processes $N\geq3$, it is possible to construct
other unravelings invariant under unitary transformation.  For
instance:
\begin{equation}
d\zeta = \frac{dW_1+ e^{i\pi/3}dW_2 + e^{-i\pi/3}dW_3}{\sqrt{3}} 
\end{equation}
where we have omitted the index $j$ and the phase factor. Since all
these unravelings have the same correlations $d\zeta^2=0$ and
$|d\zeta|^2=dt$, they are equivalent~\cite{Gardiner1} and can be replaced
by the QSD unraveling.

\section{Localization}
\label{SecLoc}

As an application of the set of continuous unraveling obtained in the
present work, one can compute the rate of self-localization of a
single environment operator $L$ for a wide open system, i.e. $H=0$,
and determine the effect of the noise correlation on the rate of
self-localization. The rate of self-localization is defined as the
rate at which the ensemble average of the quantum mean square
deviation decays \cite{P94}. It is also the ensemble average rate at
which the state vector $|\psi\>$ tends towards one of the (right-)
eigenstates of the Lindblad operator. The quantum mean square
deviation\footnote{Note that the quantum mean square deviation is not
an ensemble average.} of the operator $L$ is defined as $\sigma^2(L) =
\<L^\dagger L\>_\psi - \<L^\dagger\>_\psi\<L\>_\psi$.  More generally
the quantum covariance of two operators for the state $|\psi\>$ is
$\sigma(A,B)=\<A^\dagger B\>_\psi -\<A^\dagger\>_\psi\<B\>_\psi$
\cite{P94}. Note that the quantum covariance of $L$ with itself is
just the quantum mean square deviation $\sigma^2(L)=\sigma(L,L)$.  We
restrict our attention to a wide open system because we want to
describe the localization process, independently of the action of the
Hamiltonian. This, clearly, is only a first step towards a proper
understanding of localization which should involve Hamiltonian effects
as well.

\subsection{Hermitian environment operator} 
For a wide open system with a hermitian environment operator
$L=L^\dagger$ the state vector $|\psi\>$ evolves according to
\begin{equation}
 |d\psi\> = - \half \left( L^\dagger L +\<L^\dagger\>_\psi\< L\>_\psi 
- 2\<L^\dagger\>_\psi L \right)|\psi\> dt 
+ \left( L -\< L\>_\psi \right)|\psi\> d\zeta 
\label{EqState1L}
\end{equation}
where $d\zeta$ is a noise of the kind previously described by
(\ref{EqNoise1}) and (\ref{EqNoise2}) with an associated correlation
factor $c$ given by $d\zeta^2=c dt$. When the state evolves according
to equation (\ref{EqState1L}) we can compute the change in the
expectation value of $L$
\begin{equation}
d\<L\>_\psi = 2 \sigma^2(L) \mbox{Re}(d\zeta)
\end{equation}
and the change in the quantum mean square deviation
\begin{equation}
d\sigma^2(L) = -2\sigma^2(L)^2 \left( 1+\mbox{Re}(c) \right) dt
+2\sigma(L_\Delta L_\Delta,L) \mbox{Re}(d\zeta)
\end{equation}
where we have used the notation $L_\Delta = L-\<L\>_\psi$. The
diffusion of the expectation value $\<L\>_\psi$ and the quantum mean
square deviation is produced only by the real part of the noise term
$d\zeta$.

Taking the ensemble mean shows that the expectation value
$M\<L\>_\psi=\mbox{Tr}(\rho L)$ remains constant and the quantum mean
square deviation evolves as
\begin{equation}
M\frac{d\sigma^2(L)}{dt} = -2\left( 1+\mbox{Re}(c)\right) 
M\left(\sigma^2(L)^2\right)
\end{equation} 
The noise correlation $c$ is a characteristic signature of the chosen
unraveling.  For $c=0$, the quantum state diffusion result, giving a
minimal localization rate of 2, is recovered \cite{P94}. In this case,
as in almost all cases, the mean square deviation tends to zero, thus
the state $|\psi\>$ evolves towards one eigenstate of $L$. The real
noise (RN) unraveling $c=1$ is clearly the one which gives the highest
rate of self-localization.  As a consequence, for numerical
simulations involving an arbitrary Hamiltonian and one hermitian
environment operator, the RN unraveling should be used since it will
produce the fastest localization (for continuous unravelings).  In the
opposite case to the RN unraveling, if one uses the imaginary noise
unraveling $c=-1$, the mean localization rate vanishes and the state
does not evolve towards an eigenstate of $L$. Recovering these well
known results \cite{PB,GPPC} confirms the validity of equation
(\ref{EqState1L}).

\subsection{Non hermitian environment operator} 
We consider the case of a single non hermitian Lindblad
operator. Since this case is more difficult to treat, we restrict
ourselves to the more specific case of an annihilation operator
$L=\sqrt{\kappa}a$. As for a hermitian operator, we want to
determine which is the unraveling with the highest localization rate
and find out if there is any unraveling which does not localize.

The state vector evolves according to equation (\ref{EqState1L})
and the change in the quantum mean square deviation
$\sigma^2(a)=\<a^\dagger a\>_\psi - \< a^\dagger\>_\psi \< a\>_\psi$
is given by
\begin{eqnarray}
d\sigma^2(a) &=& -\kappa\left(\sigma^2(a)+\sigma^2(a)^2
+|\sigma(a^\dagger,a)|^2+ 2\sigma^2(a)\mbox{Re}\{ c
\sigma(a^\dagger,a)\}\right) dt \nonumber \\ && + 2\sqrt{\kappa}
\mbox{Re} \left\{\left( \sigma(a^\dagger a,a) -\<a^\dagger \>_\psi \sigma
(a^\dagger ,a) -\< a\>_\psi \sigma^2 (a)\right) d\zeta \right\}
\label{EqQcorr1}
\end{eqnarray}
which involves the quantum covariance $\sigma(a^\dagger,
a)=\<a^2\>_\psi - \< a\>_\psi^2$. The equation for $\sigma( a^\dagger,
a)$ can also be derived to give
\begin{eqnarray}
d\sigma(a^\dagger ,a) &=& -\kappa \left( (1+2\sigma^2(a)) \sigma(a^\dagger,a)
+c \sigma(a^\dagger,a)^2 +c^\ast\sigma^2(a)^2  \right) dt \nonumber \\ 
&& + 2\sqrt{\kappa}\mbox{Re}\left\{ \left( \sigma(a^2,a)-2\<a^\dagger\>_\psi\sigma^2(a)\right) d\zeta \right\}
\label{EqQcorr2}
\end{eqnarray}
For the QSD unraveling $c=0$, the two equations (\ref{EqQcorr1}) and
(\ref{EqQcorr2}) are known to describe the localization of the state
$|\psi\>$ towards a coherent state in the case of a harmonic
oscillator. Furthermore, this localization is known to be globally
stable \cite{Z96}.

Taking the ensemble mean over equation (\ref{EqQcorr1}) removes the
noise terms but introduces statistical correlations since the
statistical mean of a product is in general different from the product
of the statistical means. Thus one cannot obtain an immediate result
for the mean rate of localization. However, one can notice that the
drift part of $d\sigma^2(a)$ can be written as a sum of positive
terms:
\begin{equation}
\sigma^2(a)+(\sigma^2(a)-|\sigma(a^\dagger,a)|)^2 
+ 2\sigma^2(a)|\sigma(a^\dagger,a)|\left( 1+\mbox{Re}\{ 
c\frac{\sigma(a^\dagger,a)}{|\sigma(a^\dagger,a)|}\}\right)
\label{Eqdrift}
\end{equation}
the third term of this expression being positive since $|c|\leq 1$.
As a consequence, the argument for global stability of coherent states,
\begin{equation}
M\frac{d\sigma^2(a)}{dt} \leq -\kappa M\sigma^2(a),
\end{equation}
is valid for all the continuous unravelings. A localization rate of
$\kappa$, associated with the exponential decay of quantum
correlations $M\sigma^2(a)\simeq \exp(-\kappa t)$, is expected to be
independent of the choice of unraveling. This result shows that, for
an annihilation operator, all unravelings localize and $\kappa$
provides a minimal bound, independent of the unraveling, for the
ensemble mean localization rate.

For a hermitian operator, the unraveling which localizes the most was
easy to find since the evolution of the quantum mean square deviation
is not coupled to any other moment. Furthermore the correlation factor
$c$ factorises, making the unraveling independent of the state. In the
present case the situation is more complex, since none of these two
simplifying conditions are satisfied.  In the case of an annihilation
operator, we adopt the following technique. Instead of considering the
localization of an arbitrary state $|\psi\>$, we restrict our
attention to squeezed states. We will show that every unraveling
(\ref{EqState1L}) with $L=\sqrt{\kappa} a$ preserves the set of
squeezed states, i.e., if the initial state is a squeezed state it
will evolve into a squeezed state. This has been shown for the QSD
unraveling in ref. \cite{Z96}. The unraveling which reduces the
squeezing most efficiently is determined. Finally, some arguments will
be given as to why this unraveling should be the one with the highest
localization rate for an arbitrary initial state.

Squeezed states are defined as the states $|\gamma_t,\alpha_t\>$ which
satisfy the relation
\begin{equation}
(a -\gamma_t a^\dagger -\alpha_t) |\gamma_t,\alpha_t\> =0
\label{EqDefSS}
\end{equation}
where $\gamma_t$ and $\alpha_t$ are complex numbers which label the
squeezed state $|\gamma_t,\alpha_t\>$ \cite{M93}. When the squeezing parameter
$\gamma_t $ vanishes, squeezed states reduce to coherent states.
Amongst the properties of squeezed states, we will use for our
present purposes only the relations between the mean square deviation
and the squeezing parameter
\begin{equation}
\sigma^2_s(a) = \frac{|\gamma_t|^2}{1-|\gamma_t|^2}= \gamma_t \sigma_s(a^\dagger,a)^\ast
\label{EqMSDSS}
\end{equation}
where the index $s$ specifies that the mean square deviation is taken
with respect to the squeezed state $|\gamma_t,\alpha_t\>$. This last
relation tells us that the mean square deviation depends only on the
squeezing parameter.

A condition to check if squeezed states are preserved can be obtained
by differentiating (\ref{EqDefSS}) \cite{HZ95,RG96}. In order to
simplify the calculation, the Stratonovich formalism is used. In this
formalism, the usual differentiation rules apply. Thus from
(\ref{EqDefSS}), a state $|\psi\>$ initially squeezed will remain
squeezed if it is possible to write
\begin{equation}
(a-\gamma_t a^\dagger -\alpha_t) |d\psi\> = (d\gamma_t a^\dagger + d\alpha_t)|\psi\>
\label{EqCondSS}
\end{equation}
where $|d\psi\>$ is to be expressed in Stratonovich form. From
equation (\ref{EqState1L}) and using the usual conversion formula $X
\circ dY = XdY +\frac{1}{2} dXdY$ between the Stratonovich and It\^o
formalism \cite{HE80}, one can express the differential increment for
the state vector $|\psi\>$ as
\begin{eqnarray}
|d\psi\> &=& -\frac{1}{2}\left\{ L^\dagger L - 2\< L^\dagger \>_\psi (L-\< L\>_\psi) \right\} |\psi\> dt \nonumber \\ 
&& -\frac{c}{2}\left\{ L^2 - \<L^2\>_\psi -2\<L\>_\psi(L-\<L\>_\psi) \right\} |\psi\> dt \nonumber \\
&& + (L-\< L \>_\psi)|\psi\> \circ d\zeta
\end{eqnarray}
Inserting this expression in the condition (\ref{EqCondSS}), one finds
not only that squeezed states are preserved but also the equations for
the squeezing parameters are
\begin{eqnarray}
d\gamma_t &=& -\kappa \gamma_t (1+c\gamma_t) dt \label{EqSqueezing} \\ 
d\alpha_t &=& -\frac{\kappa}{2} \alpha_t dt +\kappa \gamma_t
\<a^\dagger\>_s (1+c\gamma_t) dt +\sqrt{ \kappa}\gamma_t d\zeta
\end{eqnarray}
written in It\^o form. Since the evolution of the squeezing parameter
$\gamma_t$ is deterministic, it is easy to find the unraveling which
produces the fastest squeezing decay. It is given by the following
correlation factor
\begin{equation}
c = \frac{\gamma_t^\ast}{|\gamma_t|}
\label{EqBestSqueezed}
\end{equation}
Notice that this unraveling depends on the state itself. In order to
produce the maximal decay in squeezing, the noise term in the
stochastic Schr\"odinger equation has to evolve according to the
prescription given in (\ref{EqBestSqueezed}). For such an unraveling,
the norm of the correlation factor $c$ is kept constant at its maximal
value $|c|=1$ and the phase varied in time along each single
trajectory in order to minimize the squeezing.

For all the unravelings with a correlation factor having such phase
but an arbitrary norm $\{ c_r=r\gamma_t^\ast/|\gamma_t|,~~ r\in
[0,1]\}$, the localization rate can be easily computed. This set
includes QSD which as $r=0$ and the adaptive unraveling
(\ref{EqBestSqueezed}) which has $r=1$. For every unraveling in this
set, the squeezing parameter $\gamma_t$ decays as $\gamma_t\simeq
e^{-\kappa t}$ for large enough times. Using the relation
(\ref{EqMSDSS}), one obtains
\begin{equation}
\sigma^2(a)\simeq e^{-2\kappa t} ~~~~~\sigma(a^\dagger, a)\simeq e^{-\kappa t} 
~~~~~\mbox{for}~~~ t\gg \kappa t
\end{equation}
Compared to the minimal localization rate of $\kappa$, this result
shows that the unravelings which have a correlation factor $c$ with
the proper phase, i.e. all unravelings in the above mentioned set,
produce a localization rate on squeezed states which is twice the
minimal localization rate.  Furthermore, the quantum mean square
deviation $\sigma(a^\dagger, a)$ decays at the same rate as the
energy. Thus, the time needed for a squeezed state to become a
coherent state is of the same order as the dissipation time, making
the squeezed state a favourite basis for numerical simulations.

Squeezed states are frequently represented in phase space using
distribution functions. The $Q$ distribution of a squeezed state is a
Gaussian whose contour forms an ellipse. It can be easily seen that
the phase (\ref{EqBestSqueezed}) of the correlation factor $c$ is
equal to $e^{i2\phi}$ where $\phi$ is the angle between the real axis
and the major axis of the ellipse. The correlation factor $c$ is
defined, irrespective of its norm, as the square of the noise term
$d\zeta$ (see (\ref{EqNoise2})). Thus the phase of the noise term
$d\zeta$, which produces the fastest squeezing decay, is the same as
the phase of the major axis of the ellipse, this later being defined
as $e^{i\phi}$.  One can say that the unraveling varies in time in
order to distribute the noise fluctuations along the axis of largest
spread of the squeezed state.

If the system state is not a squeezed state, the previous derivation
does not apply anymore. Nevertheless, we can try to generalize the
result for an arbitrary state. Using the relation (\ref{EqMSDSS})
between squeezing parameter and mean quantum deviation, the same
unraveling can be specified as
\begin{equation}
c = \frac{\sigma(a^\dagger,a)^\ast}{|\sigma(a^\dagger,a)|} 
\label{EqNewUnravel}
\end{equation}
This expression allows us to generalize the criteria obtained for a
squeezed state to an arbitrary state. This choice corresponds to
fluctuations distributed along the major axis of the ellipse which
will give the best approximation to the state distribution in phase
space.  Another interesting aspect of the unraveling
(\ref{EqNewUnravel}) is that it is also the one which maximizes the
drift term (\ref{Eqdrift}) in the $d\sigma^2(a)$ equation. This is not
a sufficient condition to insure this unraveling localizes the most
since this condition neglects statistical correlations. As soon as
statistical correlations can be neglected, this unraveling will
produce the highest localization rate.

\begin{figure}[htbp]
\centerline{\psfig{file=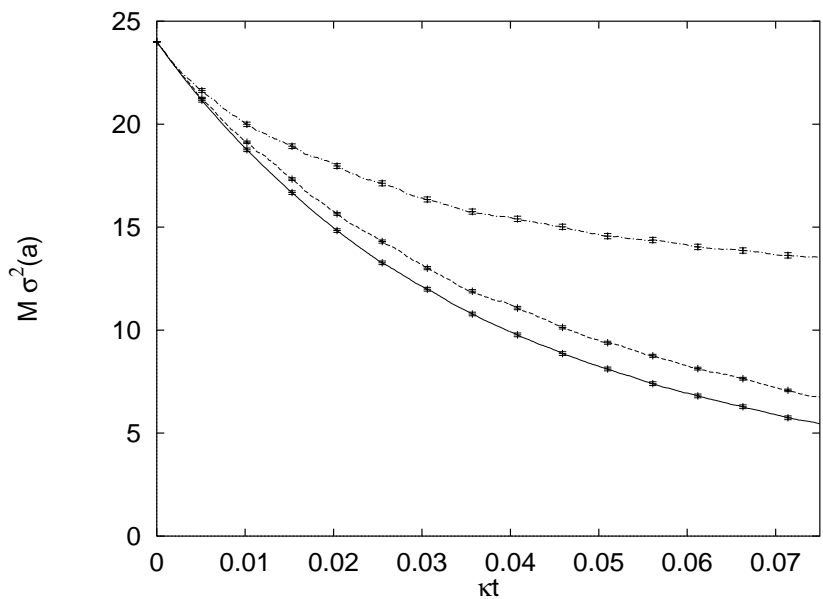,width=4.5in}}
\caption{\label{fig1} Ensemble average of the quantum mean square
deviation $\sigma^2(a)$ showing the short time scale localization. The
initial state is the Fock state $| 24\>$. Each curve represent a
different unraveling: the unraveling
(\protect\ref{EqNewUnravel})(\full), QSD (\dashed) and Real Noise
(\chain). The ensemble average is computed using 1000
trajectories. The errors bars indicate the 95\%-confidence intervals.}
\end{figure}

\begin{figure}[htbp]
\centerline{\psfig{file=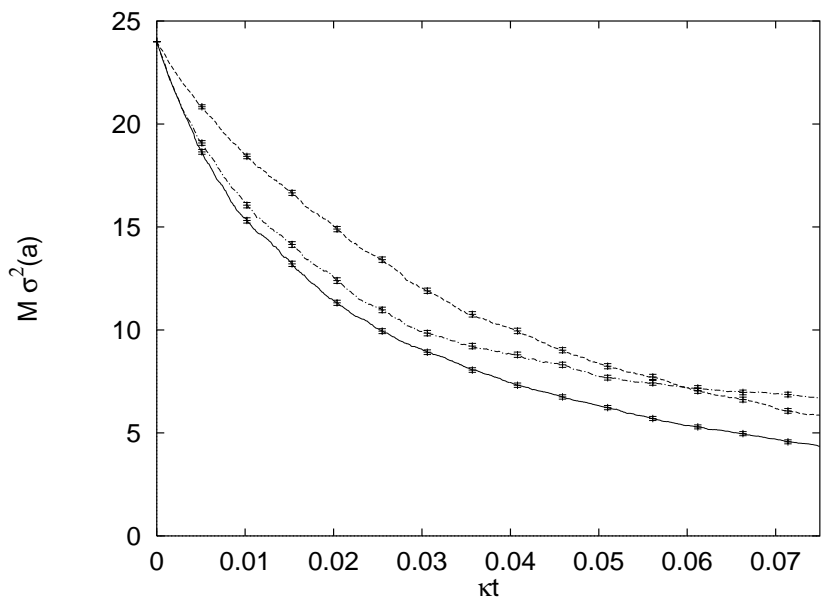,width=4.5in}}
\caption {\label{fig2} As figure \protect\ref{fig1}, but with the initial
state in a superposition of two Fock states $2^{-1/2}(|23\>+|25\>)$.}
\end{figure}

\begin{figure}[htbp]
\centerline{\psfig{file=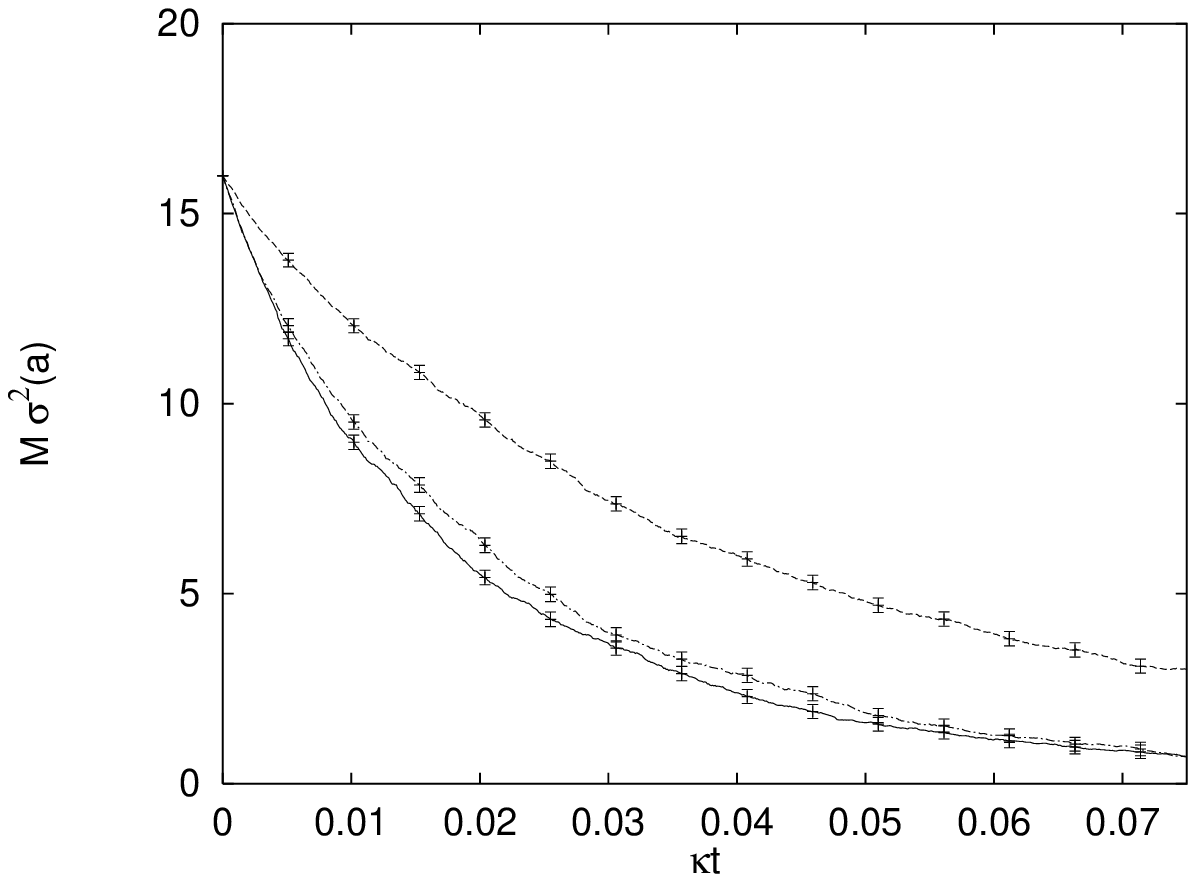,width=4.5in}}
\caption {\label{fig3} As figure \protect\ref{fig1}, but with the initial
state in a superposition of two coherent states $2^{-1/2}(|\alpha\>
+|-\alpha\>)$ with $\alpha=4$.}
\end{figure}

These two arguments show that for distributions which can be well
approximated by ellipses, the unraveling (\ref{EqBestSqueezed}) will
give the highest localization rate. What happens with other states? To
answer this question, numerical simulations have been employed. Using
three different initial states, and three different unravelings, the
evolution in time of the quantum mean square deviation $\sigma^2(a)$
has been computed. The ensemble average $M\sigma^2(a)$ taken over 1000
trajectories is represented in figures \ref{fig1}-\ref{fig3}. The
initial states chosen are: a Fock state $|24\>$, a superposition of
Fock states $(|23\>+|25\>)/\sqrt{2}$ and a superposition of coherent
states $(|\alpha\>+|-\alpha\>)/\sqrt{2}$ with $\alpha=4$. The
unravelings chosen are: the unraveling with the proper phase
(\ref{EqNewUnravel}), QSD with $c=0$ and real noise with $c=1$.

All our simulations confirm that such quantum states are very unstable
under the effect of dissipation \cite{GK94b}, this being independent
of the chosen unraveling.  After a small time $\kappa t < 1$, the
state becomes a squeezed state to a very good approximation. During
the transition from an arbitrary initial state to an almost squeezed
state, the rate of localization is very high.

Once the state has become a squeezed state, the mean rate of
localization is approximately constant for all the unravelings. The
localization rate always lies in between the minimal rate $\kappa$ and
the squeezed state rate $2\kappa$ (see figure 4). The latter occurs
only for the set of unravelings with the appropriate phase.  For the
unravelings with a different choice of phase, the term $c\gamma_t$ in
equation (\ref{EqSqueezing}) will give a negative contribution and
produce a lower localization rate. In all numerical simulations, the
new unraveling produced the highest localization.

\begin{figure}[htbp]
\centerline{\psfig{file=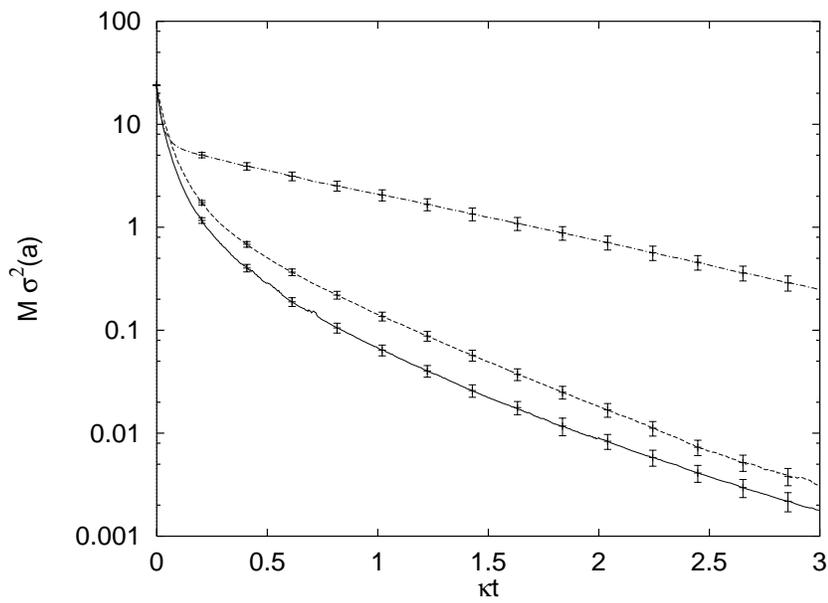,width=4.5in}}
\caption {\label{fig4} As figure \protect\ref{fig2}, but for a longer time
scale.}
\end{figure}

Finally, one can try to generalize the previous result to an arbitrary
Lindblad operator. In this case, the unraveling which localize the
most will be the one with a correlation factor given by
\begin{equation}
c = \frac{\sigma(L^\dagger,L)^\ast}{|\sigma(L^\dagger,L)|} 
\label{EqUnravel}
\end{equation}
For a hermitian operator $q$, this expression reproduces the result
previously obtained, namely
\begin{equation}
c = \frac{\sigma(q^\dagger,q)^\ast}{|\sigma(q^\dagger,q)|}
  = \frac{\sigma^2(q)}{|\sigma^2(q)|}=1
\end{equation}
for the real noise unraveling. For the annihilation operator $L=a$
equation (\ref{EqUnravel}) is equivalent to (\ref{EqNewUnravel}).

\subsection{Numerical simulations} 

Instead of trying to justify further the unraveling (\ref{EqUnravel})
as the most localizing one for an arbitrary Lindblad operator we look
for an estimate of the improvement produced by using such an
unraveling in numerical simulations.  The test consists of computing
the quantum mean square deviation $\sigma^2(a)$ along a quantum
trajectory with both the unraveling (\ref{EqNewUnravel}) and the QSD
unraveling. The ensemble average $M\sigma^2(a)$ which gives a measure
of the size of the wave packet as well as the variance
$\mbox{Var}\{\sigma^2(a)\} = M\left(\sigma^2(a)-M\sigma^2(a)\right)^2$
which measures the fluctuations of the wave packet size are then
compared between the two unravelings. The quantity $\sigma^2(a)$ is
taken here as a measure of the size of the wave packet since it
corresponds formally to the excitation number of the state displaced
at the origin \cite{BGOR97}. For a number states basis, this measure
is proportional to the number of basis state needed to represent the
state $|\psi\>$. The system we used for this test is a kicked
anharmonic oscillator
\begin{equation}
H = i \hbar \beta(t) (a^\dagger - a) +\frac{1}{2}\hbar \chi a^{\dagger 2} a^2
\end{equation}
subject to dissipation $L=\sqrt{\kappa} a$. The driving $\beta(t)$ is
a periodic sequence of rectangular pulses of height $\beta_0$, length
$\tau_1$ and separation $\tau_2$. In this case the Hamiltonian can
also play an important role in the localization process.  This system
has been considered in previous publications \cite{SR94,GR95,RG96} as
a simple example of a chaotic system. In conjunction with this system,
the scaling transformations: $\tilde{t}=\lambda t$, $\tilde{\kappa} =
\kappa/\lambda$, $\tilde{\beta_0} = \beta_0$,
$\tilde{\chi}=\chi/\lambda^3$ have been introduced. This scaling of
the parameters does not affect the classical dynamics but when it is
introduced in the quantum equations of motion, it allows one to
perform the quantum-classical transition, by varying $\lambda$ from
$\lambda=1$ (quantum) to $\lambda\rightarrow\infty$ (classical). Here
we use this transformation for a different purpose. In the quantum
limit, nonlinear effects produce a strong delocalization which can not
be compensated by the localization effect of the dissipation. In the
classical limit, on the contrary, localization is expected to be
dominant. Thus varying the scaling factor allows us to tune the
relative strength of the delocalization.

\begin{figure}[htbp]
\leavevmode
\begin{center}
\leavevmode
\epsfxsize=3in
\epsfbox{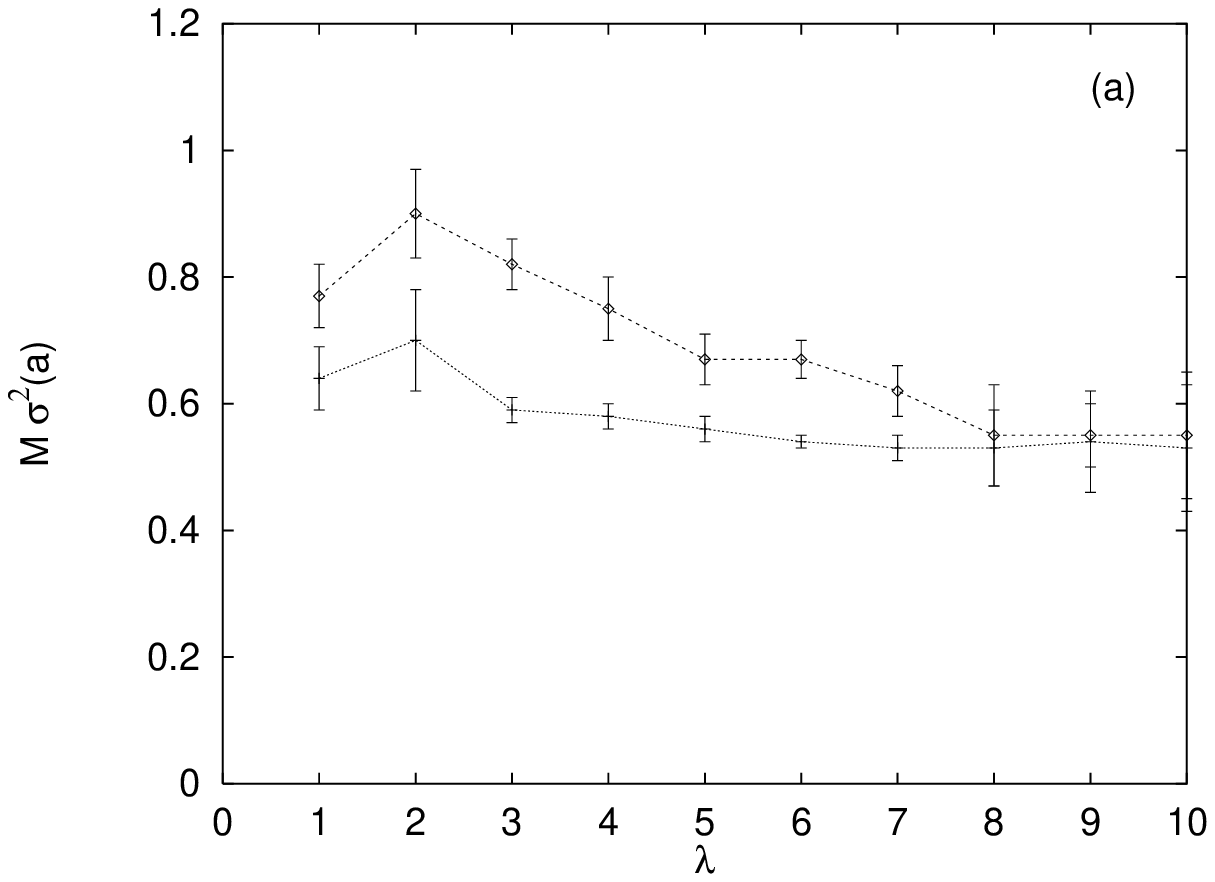}
\epsfxsize=3in
\epsfbox{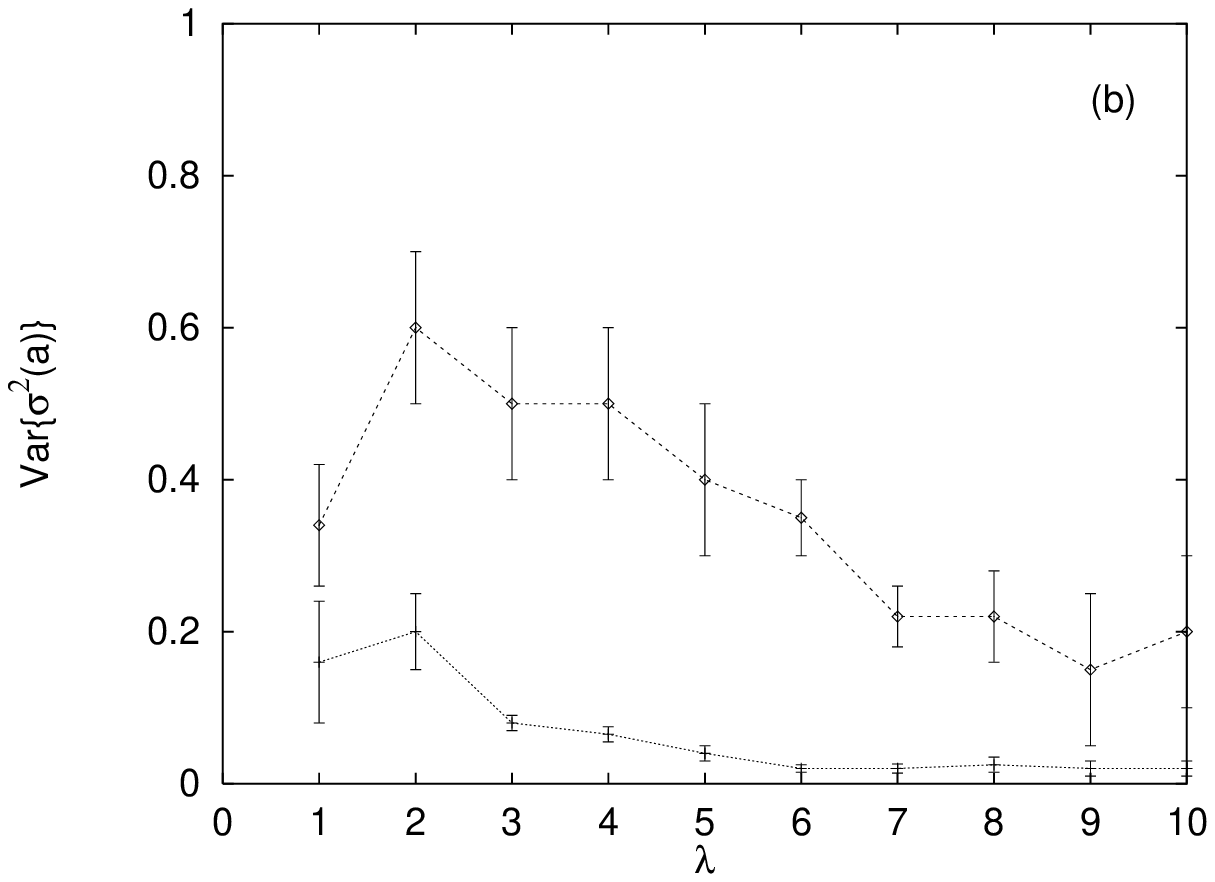}
\end{center}
\caption{\label{fig5} Ensemble average (a) and variance (b) of the
quantum mean square deviation $\sigma^2(a)$ versus the scaling
parameter $\lambda$ for the kicked anharmonic oscillator. The upper
curve corresponds to the QSD unraveling and the lower curve to the
time dependent unraveling (\protect\ref{EqNewUnravel}). The errors bars take
into account statistical as well as numerical accuracy uncertainties.}
\end{figure}

In figure \ref{fig5}, the ensemble average and variance of the quantum
square deviation $\sigma^2(a)$ are represented for different values of
the scaling parameter $\lambda$. The values represented are stationary
results obtained by integrating the equations of motions over
typically 2500 periods and taking the mean over a single
trajectory. Such a long integration in time is necessary in order to
obtain a proper average over the strange attractor of the chaotic
system. The system parameters are set to the following values
$\chi=1$, $\beta_0=2$, $\kappa=0.5$, $\tau_1=0.98$ and $\tau_2=1$ for
which the classical system is known to be chaotic. The precision of
the numerical results is estimated by repeating the calculations for
different numerical parameters such as the time step size.

Figure \ref{fig5}(a) shows for both unravelings a slowly decaying
ensemble average $M\sigma^2(a)$ for an increasing value of the scaling
$\lambda$. Notice that the amplitude of motion rescales as $\lambda$
thus the ratio $\sigma^2(a)/M\<a^\dagger a\>$ tends towards zero when
$\lambda \rightarrow\infty$, providing a numerical justification for
the emergence of the classical attractor. Furthermore, fig
\ref{fig5}(a) shows that the unraveling (\ref{EqNewUnravel}) reduces,
compared to the QSD unraveling, the stationary value of the mean size
of the wave packet. The reduction can be up to 20\% depending on the
scale parameter $\lambda$, the largest reduction being achieved in the
quantum regime.

More important is the reduction of the size of the fluctuations shown
in fig \ref{fig5}(b). The picture suggested is that each time the wave
packet deviates from a coherent state the QSD unraveling tends to
restore the shape by applying a homogeneous noise, while the
unraveling (\ref{EqNewUnravel}) adapts by applying a non-homogeneous
noise in the direction of the largest deviation. This adaptability
does not produce an important reduction of the wave packet size but
can stabilize the wave packet more efficiently as compared to QSD.

\section{Discussion}
\label{SecDiscussion}

We have introduced the set of continuous unravelings which recovers in
mean the master equation in Lindblad form and preserves the norm of
the state vector. The quantum state diffusion unraveling is a member
of this set, being the simplest which preserves the same invariance
properties under unitary transformations as the density matrix. We
have seen that each single unraveling can be specified very simply by
the choice of the noise correlations thus providing a natural
classification.  For theoretical purposes, it is useful to work with
the full set of continuous unravelings since it allows one to study
how quantities which depend on the choice of the unraveling are
sensitive to this choice.

As a first application, we have studied the localization properties
when only a single Lindblad operator is present. In the case of a
hermitian operator, the highest localization rate of the real noise
unraveling as well as the absence of localization of the imaginary
noise unraveling have been recovered and explained in a consistent
way. For a non hermitian operator, namely the annihilation operator, a
new time dependent unraveling has been introduced. It is shown
analytically that this unraveling provides the highest localization
rate for squeezed states and numerically that this property is also
valid for more complex quantum states. This unraveling maximizes the
localization by continuously adjusting the phase noise according to
the shape of the wave packet. This study provides a better
understanding of the localization. For instance, the QSD unraveling is
known to have good localization properties due to its invariance
corresponding, in some sense, to a homogeneous distribution of
noise. We have seen that the localization rate can be increased by
maximizing the norm of the noise correlation factor and adjusting
continuously its phase, this last constraint leading to the
introduction of a time dependent unraveling.

Since the new unraveling increases localization it is a good candidate
for numerical simulations of quantum trajectories and for the solution
of the master equation. A numerical comparison of the wave packet size
and fluctuations between QSD and the new unraveling shows that the new
unraveling performs better than QSD by stabilizing the size of the
wave packet.

In connection with the study of the quantum-classical transition, a
recent work by Brun \etal \cite{BGOR97} has shown that the Quantum
Jump unraveling tends to a continuous unraveling. It can be easily
seen that this unraveling is a member of the set introduced in the
present paper. We have shown that for a simple quantum system subject
to dissipation all members of the set of continuous unravelings
localize with a minimal rate given by the dissipation rate, making
localization a general property valid for all unravelings instead of
only some particular ones.

\ack 

We thank Gernot Alber, Nicolas Gisin, Ian Percival, R\"udiger
Schack and Walter Strunz for stimulating discussions. We acknowledge
financial support from the EU under its Human Capital and Mobility
Programme.

\appendix
\section{Properties of the noise correlations}
In the case of a linear combination of two Wiener process $N=2$, the
noise term $d\zeta$ is specified by the two complex numbers $\alpha_1$
and $\alpha_2$ which we write as $\alpha_1=\rho_1e^{i\phi_1}$ and
$\alpha_2=\rho_2e^{i\phi_2}$. The noise correlation factor becomes
$$
c= \frac{\sum_n \alpha_n^2}{\sum_n |\alpha_n|^2} = \frac{\rho_1^2
e^{2i\phi_1}+\rho_2^2e^{2i\phi_2}}{\rho_1^2+\rho_2^2}
$$
Using $R=\rho_2/\rho_1$ and $\theta = 2(\phi_2-\phi_1)$, this complex
number can be rewritten as
$$
c = e^{2i\phi_1} \frac{1+R^2e^{i\theta}}{1+R^2}
$$
If $\phi_1=0$ and $R$ are kept constant and $\theta$ is varied, the
denominator will give a circle in the complex plane, centered at (1,0)
and of radius $R^2$. The denominator will restrict the circle to be
inside the unit disk, centered at the origin. The phase $\phi_1$
produces only a rotation around the origin.  Thus the number $c$ can
take any value inside the unit circle only: $|c|\leq 1$. Furthermore,
this result can be generalized without difficulty for an arbitrary
number $N$ of noise terms.

\end{document}